\documentclass[journal]{IEEEtran}

\usepackage{cite,url,subfigure,epsfig,graphicx,wrapfig,psfrag} 
\usepackage{amsmath}
\usepackage{multirow}
\usepackage{amssymb}
\usepackage{algorithmic,algorithm}

\begin{document}
%
\title{Clustering-based Multicast Scheme for UAV Networks}
\author{Hao~Song, Lingjia~Liu, Bodong~Shang, Scott~Pudlewski, and Elizabeth Serena~Bentley

\thanks{H. Song is with the Next Generation and Standards (NGS) group, Intel Corporation, Santa Clara, CA 95054, USA (hao.song@intel.com).
L. Liu and B. Shang are with Bradley Department of Electrical and Computer Engineering, Virginia Tech, Blacksburg, 24060, USA. 
S. Pudlewski is with the Georgia Tech Research Institute, Atlanta,
GA 30318 USA.
E. S. Bentley is with the Air Force Research Laboratory
Information Directorate, Rome, NY 13441 USA.
Approved for Public Release; distribution unlimited 88ABW-2020-2126.
The corresponding author is L. Liu (ljliu@ieee.org).}

}

\maketitle

\begin{abstract}
 When an unmanned aerial vehicle (UAV) network is utilized as an aerial small base station (BS), like a relay deployed far away from macro BSs, existing multicast methods based on acknowledgement (ACK) feedback and retransmissions may encounter severe delay and signaling overhead due to hostile wireless environments caused by a long-distance propagation and numerous UAVs. In this paper, a novel multicast scheme is designed for UAV networks serving as an aerial small BS, where a UAV experiencing a packet loss will request the packet from other UAVs in the same cluster rather than relying on retransmissions of BSs. The technical details of the introduced multicast scheme are designed with the carrier sense multiple access with collision avoidance (CSMA/CA) protocol for practicability and without loss of generality. Then, the Poisson cluster process is employed to model UAV networks to capture their dynamic network topology, based on which distance distributions are derived using tools of stochastic geometry for analytical tractability. Additionally, critical performance indicators of the designed multicast scheme are analyzed. Through extensive simulation studies, the superiority of the designed multicast scheme is demonstrated and the system design insight related to the proper number of clusters is revealed.
\end{abstract}

\section{Introduction}

\subsection{Motivations}

Unmanned aerial vehicle (UAV) networks have recently been attractive to both the military and commercial applications in the wireless communication field due to their significant advantages, such as high mobility, easy deployment, low cost, high flexibility and high scalability [1], [2]. UAVs are useful in emergency communications, supporting communications among disaster survivors, rescue teams, and the nearest available cellular infrastructures, when commercial infrastructures, such as base stations (BSs) or power systems, get crippled in the disaster [3], [4]. Also, UAV networks are able to provide wireless access services in some special scenarios, such as communication hotspots, weak signal strength areas, and sensor networks, regardless of infrastructure, terrestrial and space topology constraints [5], [6], [29].

One of main applications of UAV networks in wireless communications is utilizing them as a small aerial BS, such as a mobile aerial relay or Pico, to provide mobile access service for a small cell or area [1], [2]. In this application, UAVs need to periodically receive messages from BSs, which may be comprised of both data and control signaling. Some messages may carry common data and control signaling, which are required to be received by all UAVs in UAV networks, especially when UAVs work cooperatively as a swarm. For example, if a UAV network is employed as a relay in long-term evolution (LTE) systems, all UAVs in the network are required to receive scheduling and power control signaling carried by physical downlink control channels (PDCCHs) [7]. Besides, if a UAV network works as distributed multiple-input multiple-output (MIMO) to strengthen the received signal power and enhance capacity, all UAVs need to receive data from a BS to perform beamforming transmissions [8]. Apparently, in these cases multicast transmissions are essential and important, where macro BSs broadcast messages to UAV networks and all UAVs are expected to receive broadcasted messages.

\subsection{Related work}

In traditional multicast transmissions, a BS broadcasts packets to all desired receivers. After receiving a broadcast packet, receivers feed ACKs back to the BS, informing it of the reception of the broadcast packet. According to the ACK feedback, the BS can be made aware if all the desired receivers have already received the broadcast packet. If not, the BS needs to retransmit the broadcast packet until all desired receivers successfully receive it. Apparently, a packet may have to experience multiple re-transmissions especially under bad wireless environments, causing inefficient multicast and high latency. Hence, many research achievements have been made on improving the efficiency of multicast transmissions.

The main methods of multicast improvements include two categories, coding and resource allocation. Random network coding (RNC) was originally introduced by T. Ho et al. [9], where a user can decode original packets as long as it accumulates sufficient different versions of encoded packets [10]. In [11], it has been proven that RNC is able to enhance the efficiency and capacity of multicast transmissions. In [12], the authors introduced adaptive random network coding for multicast transmissions, where original packets are encoded based on the priority of data. Scalable video coding (SVC) is another popular coding method in video multicast transmissions, where video data are partitioned into a base layer and multiple enhancement layers [13], [14]. The base layer is the most important layer, which is fundamental for packet decoding with a basic video quality, while enhancement layers are used to improve the video quality. P. Li et al. in [13] introduced a video multicast algorithm with SVC, in which the modulation and coding scheme (MCS) of each video layer are optimized subject to the constraint of wireless resources. In [14], M. Condoluci et al. proposed a radio resource management policy using a subgrouping technique and SVC to improve the performance of multicast transmissions in LTE systems. As for resource allocation, the authors in [15] introduced a resource allocation framework, aimed at minimizing the number of broadcasted packets, while an adequate number of users receive services. In [16], a multicast video delivery scheme is proposed, where resource allocation and MCS are jointly optimized to satisfy users' requirements and enhance videos' qualities in fourth generation (4G) networks. In addition, the greedy algorithm is adopted in resource allocation of LTE systems to maximize the throughput of multicast device-to-device (D2D) transmissions [17]. In [18], the water-filling-based resource allocation algorithms are introduced for multicast throughput maximization in spectrum sharing systems.

Although the performance of multicast transmissions could be improved by the aforementioned methods, the special characteristics of UAV networks will downgrade the performance of them, which are designed based on ACK feedback and retransmissions. To be specific, considerable delay may be caused by retransmissions, especially when UAV networks work in a bad wireless environment or have numerous UAVs. For example, if UAV networks are deployed far away from a BS to strengthen the coverage of cell edges, many retransmissions may be needed to successfully deliver a packet, incurring a large delay. This is because after a long-distance propagation from the BS to UAVs, the signal received by UAVs may be very weak. Additionally, the delay issue caused by retransmissions may be deteriorated with a large amount of UAVs, as a BS has to keep retransmitting a packet until all the UAVs successfully receive it. Then, severe control message overhead may arise because of frequent ACK feedback and numerous UAVs. UAVs are required to feed back an ACK for each received packet, while the BS has to record and process enormous ACKs from all the UAVs, which may cause a colossal burden. Hence, the traditional multicast methods based on ACK feedback and retransmissions may not be applicable in UAV networks. On the other hand, those approaches using resource allocation may be also inapplicable to UAV networks, which require accurate instantaneous channel states information. The frequent channel measurement, signal processing and feedback will cause tremendous overhead for both UAVs and BSs. Additionally, a series of system procedures are needed to enable resource allocation, including computing, resource management, and signaling exchange. With a large amount of UAVs, the resource allocation process may consume considerable time, making resource allocation results outdated and unable to adapt to the latest wireless environments.

\subsection{Our work and contributions}

In this paper, a novel multicast scheme is designed for UAV networks without relying on ACK feedback, retransmissions or resource allocation. In the designed multicast scheme, the whole UAV networks will be partitioned into multiple clusters. Rather than recovering lost packets through retransmissions, a UAV will request lost packets from other UAVs in the same cluster, when packet loss occurs. A sophisticated system design has been made for the technical details of the introduced multicast scheme, enabling it to work in UAV networks configured with the carrier sense multiple access with collision avoidance (CSMA/CA) technologies, which is likely to be used by UAVs to support wireless transmissions between them. Moreover, with an appropriate system design, some technical issues of the proposed multicast scheme could be solved, such as signaling and packet storm when multiple UAVs request the same lost packet. Then, the Poisson cluster process (PCP), a more realistic way compared to the assumption of a stationary network topology, is utilized to model UAV networks to capture their dynamic network topology, where the locations of UAVs are assumed to be random and unknown [19], [20]. To develop a more tractable model for UAV networks serving as an aerial small BS, the distance distributions from a BS to a randomly chosen UAV and between two arbitrary UAVs are derived using tools of stochastic geometry. Based on those, the comprehensive performance analysis is conducted for the designed multicast scheme in terms of coverage probability, transmission success probability, delay, and area spectrum efficiency. Using the performance analysis and the corresponding simulation studies, we aim to reveal the system design insight regarding the proper number of clusters.

The reminder of the paper is organized as follows. In Section II, UAV networks working as an aerial small BS is modeled with the Poisson cluster process. The designed multicast scheme and the relevant technical details are described in Section III. In Section IV, the distance distributions are derived, based on which the performance is analytically evaluated in Section V. Finally, Section VI and Section VII show simulation studies and conclude the whole paper, respectively.

\section{System Model}

A UAV network, consisting of multiple UAVs, is considered, which serves as a small aerial BS to provide the wireless coverage for a small area as shown in Fig. 1. The UAV network periodically receive data and control signaling through the multicast of a macro BS and forward the data to users. 

\vspace{-0em}\begin{figure}[t] 
  \begin{center}
    \scalebox{0.3}[0.3]{\includegraphics{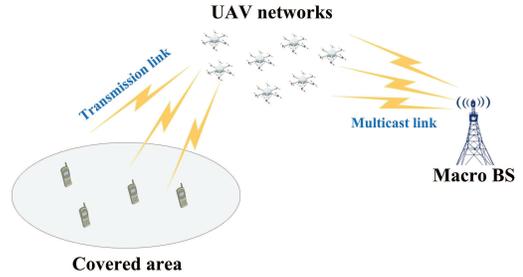}}
     \vspace{-0em}\caption{UAV networks used as a small aerial BS.}
    \label{fig:1}
  \end{center}
\vspace{-0em}\end{figure}

\subsection{Network layout}

Generally, the UAV network has a dynamic network topology due to UAVs’ high mobility. Even if a UAV network is deployed in a fixed place, for example, providing the wireless coverage for a small cell, it still has dynamic network topology. This is because UAVs may move around due to weather, like wind, or trying to avoid colliding with each other or with other obstacles. As a result, if UAVs do not have a specific flight formation, their positions will be disordered and random. To realize packet request between UAVs in the same cluster, UAVs in the same cluster should stay together and be distributed in a limited range, so that those UAVs are able to hear each other. This distribution range is defined by their cluster center and the size of the cluster. In other words, positions of UAVs in the same cluster are related to their cluster center, while UAVs are independently distributed around their cluster center in a particular range.

To capture the aforementioned features, the Poisson cluster process is applied to model UAV networks, including a parent point process and offspring point processes. To be specific, a parent point process is used to model the locations of cluster centers, which are distributed using the Poisson point process (PPP) $\mathbf{\Phi}$  with a density of $\lambda$ [21]. The randomness of UAVs’ movements will make cluster centers move in random either. Therefore, PPP is used to model the randomness of cluster centers. Then, given a cluster center, an offspring point process is utilized to model the locations of UAVs in this cluster, also referred to as cluster members. To limit UAVs in the same cluster distributed in a limited range, UAVs are independently distributed around their cluster center with the uniform distribution [19], [20]. Fig. 2 gives an example where cluster centers are distributed within a circular region with a 100 $m$ radius and the PPP density is ${{1 \times {{10}^{ - 4}}} \mathord{\left/
 {\vphantom {{1 \times {{10}^{ - 4}}} {{m^2}}}} \right.
 \kern-\nulldelimiterspace} {{m^2}}}$. Then, for each cluster, there are 10 UAVs uniformly distributed around their cluster center within a circular region with a 50 $m$ radius.

\vspace{-0em}\begin{figure}[t] 
  \begin{center}
    \scalebox{0.55}[0.55]{\includegraphics{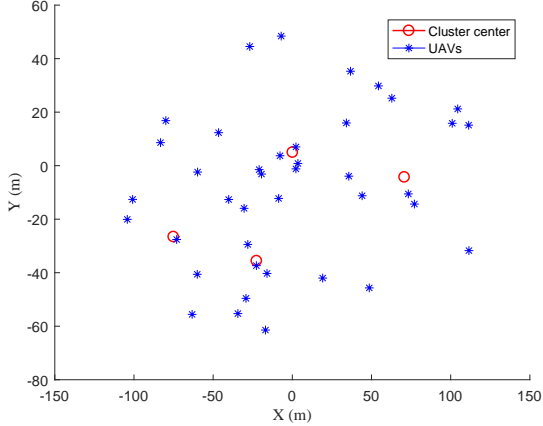}}
     \vspace{-0.0em}\caption{UAV network topology with the cluster center density of $1\times10^{-4}/{m^2}$  and 10 UAVs in each cluster.}
    \label{fig:1}
  \end{center}
\vspace{-0em}\end{figure}

\subsection{Assumptions}

Two assumptions are made in this paper based on characteristics of realistic UAV networks. First, UAVs are generally simple and low-cost devices with limited battery supply. Besides, for a long lifetime, most of the battery's energy is consumed on flying. Therefore, we assume that the transmit power of UAVs is a small value for saving energy, which is set to be 10 mW in this paper. Second, due to the simple device nature of UAVs and without loss of generality, UAVs are assumed to be configured with the CSMA/CA protocol to support wireless communications between them [22]. It would be more practical for UAVs to communicate with each other in a self-organizing manner, as wireless communications between UAVs realized by centralized control and scheduling would cause tremendous overhead for both BSs and UAVs. With the use of CSMA/CA, clear channel assessment (CCA) needs to be conducted before accessing a channel to avoid interference. 

\subsection{Channel model}

The main notations used in this paper are listed as follows. $\mathbf{\Phi}  = \left\{ {c\left| {c = 1,2, \cdot  \cdot  \cdot ,C} \right.} \right\}$ and $\mathbf{\mathbb{U}} = \left\{ {u\left| {u = 1,2, \cdot  \cdot  \cdot ,U} \right.} \right\}$ stand for the sets of clusters and UAVs, respectively. ${\mathbf{\Omega} _c} = \left\{ {{u_c}\left| {{u_c} = 1,2, \cdot  \cdot  \cdot ,{U_c}} \right.} \right\}$ represents the set of all UAVs in cluster $c$, $\bigcup\limits_{c \in \mathbf{\Phi} } {\mathbf{\Omega} _c} = \mathbb{U}$. With the CSMA/CA protocol applied in UAV networks, a UAV will access a channel only if the channel is detected to be idle. As a result, interference between UAVs could be neglected and the signal-to-noise ratio (SNR) is adopted to characterize channel states, which is dependent on many factors, including transmit power, path loss, and small-scale fading. Here, the WINNER II channel model is adopted to calculate path loss by [23]:

\vspace{-0em}\begin{equation}
\begin{array}{l}
P{L_{}} = {10^{{{ - \left( {P{L_0} + A \cdot {{\log }_{10}}\left( {d[m]} \right) + B \cdot {{\log }_{10}}\left( {{{{f_c}[GHz]} \mathord{\left/
 {\vphantom {{{f_c}[GHz]} 5}} \right.
 \kern-\nulldelimiterspace} 5}} \right)} \right)} \mathord{\left/
 {\vphantom {{ - \left( {P{L_0} + A \cdot {{\log }_{10}}\left( {d[m]} \right) + B \cdot {{\log }_{10}}\left( {{{{f_c}[GHz]} \mathord{\left/
 {\vphantom {{{f_c}[GHz]} 5}} \right.
 \kern-\nulldelimiterspace} 5}} \right)} \right)} {10}}} \right.
 \kern-\nulldelimiterspace} {10}}}},
 \end{array}
\end{equation}
where $ {PL}_0$, $d$, and $f_c$ are the path loss of a reference distance, the propagation distance, and carrier frequency, respectively. $A$ and $B$ represent the path-loss parameters with respect to distance and carrier frequency, respectively.

On the other hand, since the density of a UAV network may be high and a UAV may be surrounded by other UAVs, communications between different UAVs and between a macro BS and a UAV may experience non-link-of-sight (NLOS) transmissions and multi-path effect. Therefore, Rayleigh fading is used to model the small-scale fading. The channel gain of the link from a transmitter $i$ to a receiver $j$ is given by:

\vspace{-0.0em}\begin{equation}
\begin{array}{l}
{g_{ij}} = \sqrt {P{L_{ij}} }  \cdot {h _{ij}}
\end{array}
\end{equation}
where ${h_{ij}}$ is the small-scale fading component, obeying Rayleigh fading.
In (2), the receiver $j$ is a UAV, while the receiver $j$ could be either a UAV or a macro BS.
Accordingly, the SNR of received signals at the receiver $j$ is:

\vspace{-0.0em}\begin{equation}
\begin{array}{l}
SN{R_{ij}} = \frac{{{p_i} \cdot {{\left| {{g_{ij}}} \right|}^2}}}{{B \cdot {N_0}}} = \frac{{{p_i} \cdot P{L_{ij}} \cdot {{\left| {{h_{ij}}} \right|}^2}}}{{B \cdot {N_0}}}
\end{array}
\end{equation}
where $p_i$, $B$, and $N_0$ denote transmit power of the transmitter $i$, channel bandwidth, and noise spectral density, respectively.


\section{Multicast Scheme Based on Clustering}

\vspace{-0em}\begin{figure}[t] 
  \begin{center}
    \scalebox{0.45}[0.45]{\includegraphics{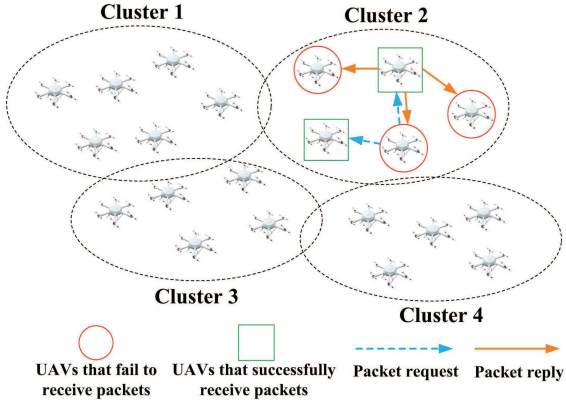}}
     \vspace{-0.0em}\caption{The designed multicast scheme in UAV networks.}
    \label{fig:1}
  \end{center}
\vspace{-0em}\end{figure}

As analyzed above, traditional multicast methods based on ACK feedback and retransmissions may encounter severe delay and control message overhead in UAV networks. To overcome these shortcomings, a multicast scheme is designed based on clustering. As shown in Fig. 3, with the designed multicast scheme, UAV networks are partitioned into multiple clusters, in which UAVs that failed to receive a packet from a BS will request the packet to other UAVs in the same cluster rather than waiting for retransmissions from the BS. By this way, retransmissions from the BS to UAVs and ACK feedback are no longer needed. Once the packet request is heard by a UAV that possesses the requested packet, it will broadcast the packet when wireless channels are available.

In a cluster, there may be multiple UAVs failing to receive a packet from the broadcast of a BS and requesting lost packets as shown in Fig. 3. Moreover, multiple UAVs that hold broadcasted packets may receive the request and attempt to convey requested packets. To avoid signaling and packet storm, a message storm avoidance policy is made herein based on the CSMA/CA protocol. To be specific, for a UAV experiencing a packet loss, it will send a packet request when wireless channels become idle. During the period of waiting for available channels, the UAV will keep sensing wireless channels. If the UAV detects that its lost packet has already been requested by another UAV or replied and sent by a UAV, responding to the request of another UAV, the UAV will receive the transmitted packet and will not send the packet request anymore. For a UAV receiving a packet request, if through spectrum sensing, it found that the packet request has been replied to by other UAVs, the UAV will not reply with the request anymore. This policy is executable owning to the property of CSMA/CA. With CSMA/CA, essential control information is comprised in the physical frame header of 802.11 packets, which is readable for all wireless devices using CSMA/CA [22], [27]. To enable our designed multicast scheme, the packet information should also be included in the header of physical frames carrying the packet request and the packet reply, so that UAVs could acquire the packet information by decoding the physical frame header.

\vspace{-0em}\begin{figure}[t] 
  \begin{center}
    \scalebox{0.5}[0.5]{\includegraphics{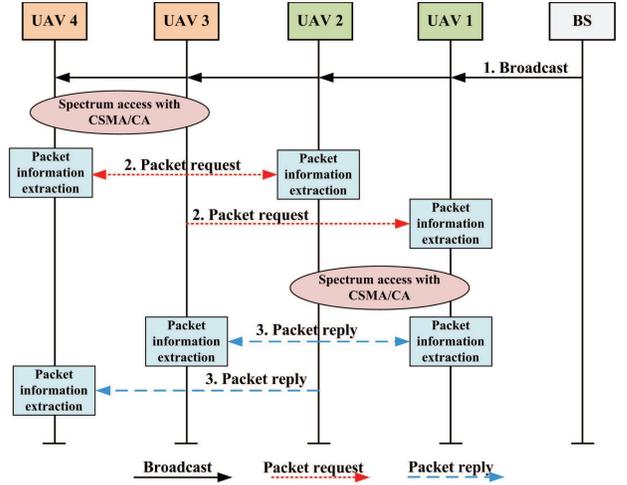}}
     \vspace{-0em}\caption{The process of the designed multicast scheme.}
    \label{fig:1}
  \end{center}
\vspace{-0.0em}\end{figure}

To elaborate the process of the designed multicast scheme, an example is given in Fig. 4, where there are four UAVs in a cluster. Assume that UAV 1 and UAV 2 successfully received broadcasted packets from a BS, while UAV 3 and UAV 4 failed, which intend to request lost packets from other UAVs in the same cluster. Assume that UAV 3 wins the spectrum access opportunity through the CSMA/CA competition and sends its packet request. UAV 1, UAV 2, and UAV 4 could receive the packet request. By extracting the packet information, UAV 4 will not send a packet request anymore, since it finds its lost packet has been requested by another UAV. On the other hand, it is assumed that UAV 2 earns the spectrum access opportunity and conveys the requested packet carried by a packet reply. After receiving the packet reply and reading packet information contained in it, UAV 1 will not reply to the request, which has been replied to by UAV 2. For UAV 3 and UAV 4, if any of them fail to receive the packet reply, they will retransmit a packet request for the lost packet.

\section{Distance Distribution}

Due to the dynamic network topology of UAV networks, the locations of UAVs are random and unknown. The distance between two randomly selected UAVs in a cluster and the distance from a BS to an UAV are random variables. In this section, the distributions of these two distances are characterized using tools of stochastic geometry, which are fundamental for channel state calculations and the performance analysis in the next section. Without loss of generality and inspired by [19], [20], [24], the distances from \emph{a typical UAV}, which is randomly chosen in a representative cluster $c \in \mathbf{\Phi}$, to a macro BS and another randomly selected UAV in the same cluster are considered. As shown in Fig. 5, it is assumed that the typical UAV is located at origin $\left( {{\text{0}},{\text{0,}}{h_2}} \right)$, while another randomly selected UAV and a macro BS are located at coordinates $\left( {{y_1},{y_2},{h_2}} \right)$ and $\left( {{z_1},{z_2},{h_1}} \right)$, respectively, for the ease of analysis. Accordingly, the distances from the macro BS to the typical UAV and between two randomly chosen UAVs could be notated by ${d_1} = \sqrt {z_1^2 + z_2^2 + \left( {{h_1} - {h_2}} \right)}$ and ${d_2} = \sqrt {y_1^2 + y_2^2}$, respectively. Moreover, assume that location of the cluster center of the representative cluster $c$ is $\left( {{x_1},{x_2},{h_2}} \right)$, in which cluster members of $c$ are uniformly distributed around the cluster center of with a radius of $r$.

\vspace{-0em}\begin{figure}[t] 
  \begin{center}
    \scalebox{0.75}[0.75]{\includegraphics{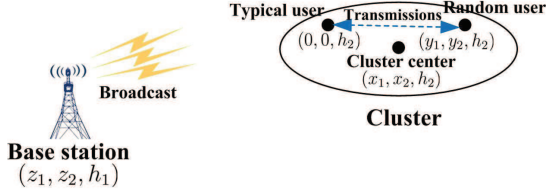}}
     \vspace{-0em}\caption{Distance model.}
    \label{fig:1}
  \end{center}
\vspace{-0em}\end{figure}

\subsection{Distance from a macro BS to a typical UAV}

Let ${\hat d_1}$ be ${\hat d_1} = \sqrt {z_1^2 + z_2^2}$ with ${d_1} = \sqrt {\hat d_1^2 + {{\left( {{h_1} - {h_2}} \right)}^2}}$. The distribution of ${\hat d_1}$ will be explored first in the X-Y plane axis, based on which distribution of $d_1$ could be obtained. Let ${\mathbf{x}_0} = \left( {{x_1},{x_2}} \right)$, $\mathbf{v} = \left( {{v_1},{v_2}} \right)$, and $\mathbf{z} = \left( {{z_1},{z_2}} \right)$ denote the vectors from the typical UAV to the cluster center, from the cluster center to the base station, and from the typical UAV to the base station, respectively. As shown in Fig. 6, $\mathbf{z} = {\mathbf{x}_0} + \mathbf{v} = \left( {{z_1},{z_2}} \right) = \left( {{x_1} + {v_1},{x_2} + {v_2}} \right)$. As ${\mathbf{x}_0}$ is a uniform random variable with the distribution of ${f_{{X_1},{X_2}}}\left( {{x_1},{x_2}} \right) = \frac{1}{{\pi {r^2}}}$, $\left\| {{\mathbf{x}_0}} \right\| \leqslant r$, the joint probability density function (PDF) of ${z_1}$ and ${z_2}$ conditioned on $\mathbf{v}$ is given by:

\vspace{-0em}\begin{align}
\begin{array}{l}
\begin{gathered}
  {f_{{Z_1},{Z_2}}}\left( {{z_1},{z_2}\left| \mathbf{v} \right.} \right) =
   \hfill \\
  {f_{{X_1},{X_2}}}\left( {{z_1} - {v_1},{z_2} - {v_2}} \right) \cdot \left| {\det \left[ {\begin{array}{*{20}{c}}
  {\frac{{\partial {x_1}}}{{\partial {z_1}}}}&{\frac{{\partial {x_1}}}{{\partial {z_2}}}} \\
  {\frac{{\partial {x_2}}}{{\partial {z_1}}}}&{\frac{{\partial {x_2}}}{{\partial {z_2}}}}
\end{array}} \right]} \right| \hfill \\
   = \frac{1}{{\pi {r^2}}},\quad \quad \quad \quad \quad  {\left( {{z_1} - {v_1}} \right)^2} + {\left( {{z_2} - {v_2}} \right)^2} \leqslant {r^2}. \hfill \\
\end{gathered}
 \end{array}
\end{align}

\vspace{-0em}\begin{figure}[t] 
  \begin{center}
    \scalebox{0.55}[0.55]{\includegraphics{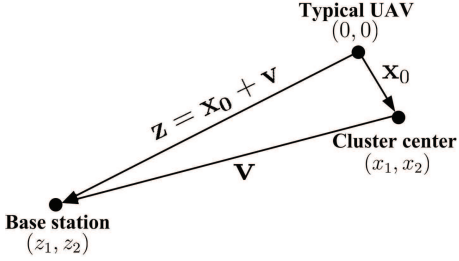}}
     \vspace{-0.0em}\caption{Model of the distance between the BS to the typical UAV.}
    \label{fig:1}
  \end{center}
\vspace{-0em}\end{figure}

With ${f_{{Z_1},{Z_2}}}\left( {{z_1},{z_2}\left| \mathbf{v} \right.} \right)$, the conditional PDF of ${\hat d_1}$ could be derived by:

\vspace{-0em}\begin{align}
\small
\begin{array}{l}
\begin{gathered}
  {f_{{{\hat D}_1}}}\left( {{{\hat d}_1}\left| \mathbf{v} \right.} \right) =
  \hfill \\
  \int_{ - {{\hat d}_1}}^{ - \frac{{\hat d_1^2 + {{\left\| \mathbf{v} \right\|}^2} - {r^2}}}{{2\left\| \mathbf{v} \right\|}}} {{f_{{Z_1},{Z_2}}}\left( {{z_1},\sqrt {\hat d_1^2 - z_1^2} \left| \mathbf{v} \right.} \right)} \left| {\frac{{\partial \sqrt {\hat d_1^2 - z_1^2} }}{{\partial {{\hat d}_1}}}} \right|
  \hfill \\
  + {f_{{Z_1},{Z_2}}}\left( {{z_1}, - \sqrt {\hat d_1^2 - z_1^2} \left| \mathbf{v} \right.} \right)\left| { - \frac{{\partial \sqrt {\hat d_1^2 - z_1^2} }}{{\partial {{\hat d}_1}}}} \right|\;\mathrm{d}{z_1} \hfill \\
  = \frac{{2{{\hat d}_1}}}{{\pi {r^2}}}\left( {\frac{\pi }{2} - \operatorname{arc} \,sin\frac{{\hat d_1^2 + {{\left\| \mathbf{v} \right\|}^2} - {r^2}}}{{2\left\| \mathbf{v} \right\|{{\hat d}_1}}}} \right),
    \hfill \\
        \quad\quad\qquad\qquad\qquad\qquad\qquad\qquad \left\| \mathbf{v} \right\| - r \leqslant {{\hat d}_1} \leqslant \left\| \mathbf{v} \right\| + r,
  \hfill \\
\end{gathered}
 \end{array}
\end{align}
where $\left\| \mathbf{v} \right\|$ represents the length of the vector $\mathbf{v}$. It is important to note that herein the the law of Cosines is used to determine the scope of $z_1$ in (5) as a function of ${\hat d}_1$, $\left\| \mathbf{v} \right\|$, and $r$. By this way, conditioning on $\mathbf{v}$ in ${f_{{{\hat D}_1}}}\left( {{{\hat d}_1}\left| \mathbf{v} \right.} \right)$ has been converted to conditioning on $\left\| \mathbf{v} \right\|$, which is a weaker condition. Accordingly, ${f_{{{\hat D}_1}}}\left( {{{\hat d}_1}\left| \mathbf{v} \right.} \right)$ can also be expressed as ${f_{{{\hat D}_1}}}\left( {{{\hat d}_1}\left| {\left\| \mathbf{v} \right\|} \right.} \right)$.

\noindent \emph{Proof}: See the detailed derivations of (5) in Appendix A.

Under ${d_1} = \sqrt {\hat d_1^2 + {{\left( {{h_1} - {h_2}} \right)}^2}}$, the distribution of $d_1$ is:

\vspace{-0em}\begin{align}
\small
\begin{array}{l}
\begin{gathered}
  {f_{{D_1}}}\left( {{d_1}\left| {\left\| \mathbf{v} \right\|} \right.} \right) = {f_{{{\hat D}_1}}}\left( {{{\hat d}_1}\left| {\left\| \mathbf{v} \right\|} \right.} \right){\left| {\frac{{\partial {{\hat d}_1}}}{{\partial {d_1}}}} \right|_{{{\hat d}_1} = \sqrt {d_1^2 - {{\left( {{h_1} - {h_2}} \right)}^2}} }}
  \hfill \\
   = \frac{{{2d_1}}}{{\pi {r^2} }} \cdot
   \left( {\frac{\pi }{2} - \operatorname{arc} \,sin\frac{{d_1^2 - {{\left( {{h_1} - {h_2}} \right)}^2} + {{\left\| \mathbf{v} \right\|}^2} - {r^2}}}{{2\left\| \mathbf{v} \right\|\sqrt {d_1^2 - {{\left( {{h_1} - {h_2}} \right)}^2}} }}} \right),
   \hfill \\
   \, \sqrt {{{\left( {\left\| \mathbf{v} \right\| - r} \right)}^2} + {{\left( {{h_1} - {h_2}} \right)}^2}}  \leqslant {d_1} \leqslant \sqrt {{{\left( {\left\| \mathbf{v} \right\| + r} \right)}^2} + {{\left( {{h_1} - {h_2}} \right)}^2}} . \hfill \\
\end{gathered}
 \end{array}
\end{align}

\subsection{Distance between two randomly selected UAVs}

Recall that ${\mathbf{x}_0}$ is the vector from a typical UAV to the cluster center and let $\mathbf{w} = \left( {{w_1},{w_2}} \right)$ be the vector the cluster center to a UAV selected at random. As shown in Fig. 7, the vector from a typical UAV to another randomly chosen UAV could be represented by $\mathbf{y} = {\mathbf{x}_0} + \mathbf{w}$ with the length ${d_2} = \left\| \mathbf{y} \right\| = \sqrt {y_1^2 + y_2^2}$. Since UAVs are uniformly distributed around the cluster center, $\mathbf{w}$ is a random variable, obeying the uniform distribution of ${f_{{W_1},{W_2}}}\left( {{w_1},{w_2}} \right) = \frac{1}{{\pi {r^2}}}$, $\left\| \mathbf{w} \right\| \leqslant r$. Correspondingly, $\mathbf{y} = \left( {{y_1},{y_2}} \right)$ is also uniformly distributed conditioned on ${\mathbf{x}_0} = \left( {{x_1},{x_2}} \right)$ with the PDF of ${f_{{Y_1},{Y_2}}}\left( {{y_1},{y_2}\left| {{\mathbf{x}_0}} \right.} \right) = \frac{1}{{\pi {r^2}}}$, ${\left( {{y_1} - {x_1}} \right)^2} + {\left( {{y_2} - {x_2}} \right)^2} \leqslant {r^2}$. The conditional PDF of ${d_2} = \sqrt {y_1^2 + y_2^2}$ could be calculated by:

\vspace{-0em}\begin{align}
\small
\begin{array}{l}
\begin{gathered}
  {f_{{D_2}}}\left( {{d_2}\left| {{\mathbf{x}_0}} \right.} \right) =
   \hfill \\
  \int_{\frac{{d_2^2 + {{\left\| {{\mathbf{x}_0}} \right\|}^2} - {r^2}}}{{2\left\| {{\mathbf{x}_0}} \right\|}}}^{{d_2}} {{f_{{Y_1},{Y_2}}}\left( {{y_1},\sqrt {d_2^2 - y_1^2} \left| {{\mathbf{x}_0}} \right.} \right)} \left| {\frac{{\partial \sqrt {d_2^2 - y_1^2} }}{{\partial {d_2}}}} \right|
  \hfill \\
  + {f_{{Y_1},{Y_2}}}\left( {{y_1}, - \sqrt {d_2^2 - y_1^2} \left| {{\mathbf{x}_0}} \right.} \right)\left| { - \frac{{\partial \sqrt {d_2^2 - y_1^2} }}{{\partial {d_2}}}} \right|\;\mathrm{d}{y_1}
  \hfill \\
   = \frac{{2{d_2}}}{{\pi {r^2}}}\left( {\frac{\pi }{2} - \operatorname{arc} \,sin\frac{{d_2^2 + {{\left\| {{\mathbf{x}_0}} \right\|}^2} - {r^2}}}{{2\left\| {{\mathbf{x}_0}} \right\|{d_2}}}} \right),\;\; 0 \leqslant {d_2} \leqslant r + \left\| {{\mathbf{x}_0}} \right\|,
   \hfill \\
\end{gathered}
 \end{array}
\end{align}
where $\left\| {{\mathbf{x}_0}} \right\|$ represents the length of the vector ${\mathbf{x}_0}$. Instead of conditioning on ${\mathbf{x}_0}$, the PDF of $d_2$ is relaxed to condition on $\left\| {{\mathbf{x}_0}} \right\|$, which is easier to be de-conditioned. As a result, ${f_{{D_2}}}\left( {{d_2}\left| {\left\| {{\mathbf{x}_0}} \right\|} \right.} \right){\text{ = }}{f_{{D_2}}}\left( {{d_2}\left| {{\mathbf{x}_0}} \right.} \right)$ suffices.

\noindent \emph{Proof}: See the detailed derivations of (7) in Appendix B.

\vspace{-0em}\begin{figure}[t] 
  \begin{center}
    \scalebox{0.55}[0.55]{\includegraphics{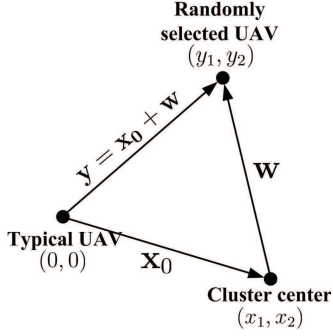}}
     \vspace{-0.0em}\caption{Model of the distance between two random chosen UAVs.}
    \label{fig:1}
  \end{center}
\vspace{-0em}\end{figure}

\section{Performance Analysis}

A UAV experiencing a packet loss could retrieve the lost packet from other UAVs in the same cluster, only if two essential conditions are satisfied: i) at least one UAV in the same cluster has received the packet; ii) the successful wireless transmissions between these two UAVs. In this paper, these two conditions are characterized by coverage probabilities and transmission success probabilities, respectively.

\subsection{Coverage probability}

According to [19], [25], the coverage probability is formally defined as the probability that the SNR of signals received by the typical UAV, which is selected at random, exceeds a pre-determined threshold ${\Gamma}$ for successful demodulation and decoding. In other words, the coverage probability could measure the possibility that a randomly chosen UAV could successfully receive packets from a BS, which can be obtained using the following:

\vspace{-0em}\begin{align}
\footnotesize
\begin{array}{l}
\begin{gathered}
  {\mathrm{P}_{cov}}\mathop  = \limits^{(a)} \int_{ - \infty }^\infty  {\mathbb{P}\left[ {SNR\left( {{D_1}} \right) > {\Gamma}\left| {{D_1}} \right.} \right]}  \cdot {f_{{D_1}}}\left( {{d_1}} \right)\;\mathrm{d}{d_1}
  \hfill \\
   = {\mathbb{E}_{{D_1}}}\left[ {\mathbb{P}\left[ {SNR\left( {{D_1}} \right) > {\Gamma}\left| {{D_1}} \right.} \right]} \right]
   \hfill \\
   = {\mathbb{E}_{{D_1}}}\left[ {\mathbb{P}\left[ {\frac{{{p_{BS}} \cdot PL\left( {{d_1}} \right) \cdot {{\left| h \right|}^2}}}{{B \cdot {N_0}}} > {\Gamma}\left| {{D_1}} \right.} \right]} \right]
    \hfill \\
   = {\mathbb{E}_{{D_1}}}\left[ {\mathbb{P}\left[ {{{\left| h \right|}^2} > \frac{{{\Gamma} \cdot B \cdot {N_0}}}{{{p_{BS}} \cdot PL\left( {{d_1}} \right)}}\left| {{D_1}} \right.} \right]} \right]
   \hfill \\
  \mathop  = \limits^{(b)} {\mathbb{E}_{{D_1}}}\left[ {exp\left( { - \frac{{{\Gamma} \cdot B \cdot {N_0}}}{{{p_{BS}} \cdot PL\left( {{d_1}} \right)}}} \right)\left| {{D_1}} \right.} \right]
  \hfill \\
   = \int_{\sqrt {{{\left( {\left\| \mathbf{v} \right\| - r} \right)}^2} + {{\left( {{h_1} - {h_2}} \right)}^2}} }^{\sqrt {{{\left( {\left\| \mathbf{v} \right\| + r} \right)}^2} + {{\left( {{h_1} - {h_2}} \right)}^2}} } {exp\left( { - \frac{{{\Gamma} \cdot B \cdot {N_0}}}{{{p_{BS}} \cdot PL\left( {{d_1}} \right)}}} \right)
   \cdot {f_{{D_1}}}\left( {{d_1}\left| {\left\| \mathbf{v} \right\|} \right.} \right)} \;\mathrm{d}{d_1}
    \hfill \\
\end{gathered}
 \end{array}
\end{align}
where $\left( a \right)$ follows the total probability theorem and $\left( b \right)$ follows the fact that the channel gain of Rayleigh fading channels is exponentially distributed. In addition, ${p_{BS}}$ is the transmit power of the BS in broadcasting. $PL\left( {{d_1}} \right)$ and $SNR\left( {{D_1}} \right)$ denote the path loss of the link from the BS to the typical UAV and the corresponding SNR with respect to the distance between them, respectively.

\subsection{Transmission success probability}

The transmission success probability refers to the probability of successful transmissions between two arbitrarily chosen UAVs in a cluster. Similar to the coverage probability, it is assumed that two UAVs are able to effectively communicate with each other, only if the SNR of received signals is greater than a threshold ${\Gamma}$ [26]. Therefore, the transmission success probability can be mathematically expressed as:

\vspace{-0em}\begin{align}
\footnotesize
\begin{array}{l}
\begin{gathered}
  {\mathrm{P}_{suc}} = \int_{ - \infty }^\infty  {\mathbb{P}\left[ {SNR\left( {{D_2}} \right) > {\Gamma}\left| {{D_2}} \right.} \right]}  \cdot {f_{{D_2}}}\left( {{d_2}} \right)\;\mathrm{d}{d_2}
   \hfill \\
   = {\mathbb{E}_{{D_2}}}\left[ {\mathbb{P}\left[ {SNR\left( {{D_2}} \right) > {\Gamma}\left| {{D_2}} \right.} \right]} \right]
   \hfill \\
   = {\mathbb{E}_{{D_2}}}\left[ {exp\left( { - \frac{{{\Gamma} \cdot B \cdot {N_0}}}{{{p_{TX}} \cdot PL\left( {{d_2}} \right)}}} \right)\left| {{D_2}} \right.} \right]
   \hfill \\
  \mathop  = \limits^{(c)} \int_0^r {\int_{0}^{r + \left\| {{\mathbf{x}_0}} \right\|} {exp\left( { - \frac{{{\Gamma} \cdot B \cdot {N_0}}}{{{p_{TX}} \cdot PL\left( {{d_2}} \right)}}} \right) \cdot } } {f_{{D_2}}}\left( {{d_2}\left| a \right.} \right) \cdot {f_A}\left( a \right)\;\mathrm{d}{d_2}\mathrm{d}a \hfill \\
\end{gathered}
 \end{array}
\end{align}
where $\left( c \right)$ represents $a = \left\| {{\mathbf{x}_0}} \right\| = \sqrt {x_1^2 + x_2^2}$. The PDF of the random variable $A$ is ${f_A}\left( a \right) = \frac{{2a}}{{{r^2}}}$, $0 \leqslant a \leqslant r$, which is used to decondition over $\left\| {{\mathbf{x}_0}} \right\|$. ${p_{TX}}$ denotes the transmit power of the transmitter UAV.

\noindent \emph{Proof}: See the derivations of ${f_A}\left( a \right)$ in Appendix C.

\subsection{Performance analysis of the designed multicast scheme}

As analyzed above, a UAV with a packet loss could request and attain the lost packet from other UAVs, only if at least one UAV in the same cluster received the packet, meanwhile UAVs are capable of effectively communicating with each other. Thus, assuming that the density of UAVs (offsprings) is ${\lambda _{off}}$, the probability that a UAV in cluster $c$ successfully obtains the lost packet through request, also referred to as request success probability, can be computed by:

\vspace{-0em}\begin{align}
\begin{array}{l}
\mathrm{P}_{req}^c = \left[ {1 - {{\left( {1 - \mathrm{P}_{cov}^c} \right)}^{\left\lfloor {{\lambda _{off}}\pi {r^2}} \right\rfloor }}} \right] \cdot {\mathrm{P}_{suc}},
 \end{array}
\end{align}
where $\left\lfloor  \bullet  \right\rfloor$ denotes the nearest integer of $ \bullet $  toward zero and $\mathrm{P}_{cov}^c$ stands for the coverage probability of cluster $c$. Notably, in each cluster, UAVs (offspring members) have the identical distribution, which are uniformly distributed around their cluster center (parent point) with a radius of $r$, resulting in the same distribution of the distance between two different UAVs. Hence, the transmission success probabilities ${\mathrm{P}_{suc}}$ of different clusters are the same, regardless of clusters. On the contrary, different clusters may have different coverage probabilities, which depend on the distance between the cluster center and the BS, $\left\| \mathbf{v} \right\|$. The distances $\left\| \mathbf{v} \right\|$ of different clusters may be varying, causing different coverage probabilities.

According to the definition of the coverage probability [19], [25], it can also reflect the probability that an arbitrarily chosen UAV can receive a packet in the broadcast of a BS. Let $L$ be the time length of a packet. If a UAV is successful in receiving the packet during broadcasting, the multicast delay would be $L$. On the other hand, if a UAV failed, the corresponding multicast delay would be $L + \frac{{L + {t_{req}}}}{{{P_{suc}}}}$, where the first term $L$ is the time length of broadcasting, while the second term $\frac{{L + {t_{req}}}}{{{P_{suc}}}}$ includes the time of packet transmissions and request signaling transmissions between UAVs. Multiple transmissions between UAVs may be needed until the packet finally reaches the UAV requesting it, since the transmit power of UAVs is generally too low to guarantee the requested packet successfully received in one transmission. Therefore, the average multicast delay of a packet in cluster $c$ could be expressed as:

\vspace{-0em}\begin{align}
\begin{array}{l}
\mathrm{Delay}{_{aver}} = \mathrm{P}_{cov}^c \cdot L + \left( {1 - \mathrm{P}_{cov}^c} \right) \cdot \left( {L + \frac{{L + {t_{req}}}}{{{\mathrm{P}_{suc}}}}} \right),
 \end{array}
\end{align}

The area spectral efficiency is formally defined as bits per second per Hertz per area unit, which is widely utilized in performance analysis of a network with a dynamic network topology [19]. For an area with the UAV (offspring) density of ${\lambda _{off}}$, the area spectral efficiency is $\mathrm{ASE} = {\lambda _{off}} \cdot {\log _2}\left( {1 + \Gamma } \right)$.
It is worth noting that the value of $\Gamma$ is determined corresponding to the types of modulation and coding. Given a type of modulation and coding, if the SNR is greater than the corresponding threshold $\Gamma$, transmitted signals in a packet can be demodulated and decoded. Similar to the analysis of the multicast delay, the area spectral efficiency varies in receiving packets through broadcasting and through transmissions between UAVs. The average area spectral efficiency of clusters $c$ could be calculated as:

\vspace{-0em}\begin{align}
\begin{array}{l}
{\mathrm{ASE}_{aver}} = \mathrm{P}_{cov}^c \cdot {\lambda _{off}} \cdot {\log _2}\left( {1 + \Gamma } \right)
 \hfill \\
 + \left( {1 - \mathrm{P}_{cov}^c} \right) \cdot {\mathrm{P}_{suc}} \cdot {\lambda _{off}} \cdot {\log _2}\left( {1 + \Gamma } \right)
 \end{array}
\end{align}

As each cluster has the identical UAV distribution, different clusters share the same ${\lambda _{off}}$.

\section{Simulation Results and Analysis}

Through simulation studies, the system design insight and the performance of the proposed multicast scheme are investigated based on the performance analysis in Section V.

\subsection{Simulation setup}

\begin{table} 
\scriptsize
\newcommand{\tabincell}[2]{\begin{tabular}{@{}#1@{}}#2\end{tabular}}
  \centering
  \caption{Simulation parameters [23], [28].}
  \vspace{-0.0em}\begin{tabular}{|c|c|}\hline
  {\textbf{Parameters}}&{\textbf{Values}}\\\hline

\tabincell{c}{Transmit power of BS} & \tabincell{c}{1000 $mW$} \\\hline

\tabincell{c}{Transmit power of UAV} & \tabincell{c}{10 $mW$} \\\hline

\tabincell{c}{Spectrum bandwidth from BS to UAV} & \tabincell{c}{20 $MHz$} \\\hline

\tabincell{c}{Spectrum bandwidth between two UAVs} & \tabincell{c}{20 $MHz$} \\\hline

\tabincell{c}{Carrier frequency from BS to UAV} & \tabincell{c}{2 $GHz$} \\\hline

\tabincell{c}{Carrier frequency between two UAVs} & \tabincell{c}{5.8 $GHz$} \\\hline

\tabincell{c}{Path loss from BS to UAV} & \tabincell{c}{$39 + 26 \cdot {\log _{10}}(d[m])$ \\$+ 20 \cdot {\log _{10}}({{{f_c}[GHz]} \mathord{\left/
 {\vphantom {{{f_c}[GHz]} 5}} \right.
 \kern-\nulldelimiterspace} 5})$} \\\hline

\tabincell{c}{Path loss between two UAVs} & \tabincell{c}{$41 + 22.7 \cdot {\log _{10}}(d[m])$ \\$+ 20 \cdot {\log _{10}}({{{f_c}[GHz]} \mathord{\left/
 {\vphantom {{{f_c}[GHz]} 5}} \right.
 \kern-\nulldelimiterspace} 5})$} \\\hline

\tabincell{c}{Noise spectral density $N_0$} & \tabincell{c}{-174 $dBm/Hz$} \\\hline

\tabincell{c}{SNR threshold ${\Gamma}$} & \tabincell{c}{20} \\\hline

\tabincell{c}{Height of BS ${h_1}$} & \tabincell{c}{10 $m$} \\\hline

\tabincell{c}{Height of UAV networks ${h_2}$} & \tabincell{c}{20 $m$} \\\hline

\tabincell{c}{Packet time length ${L}$} & \tabincell{c}{10 $ms$} \\\hline
\end{tabular}
\vspace{-0em}\end{table}

In the simulation, a UAV network consisting of 50 UAVs is considered, which is partitioned into multiple clusters. The Poisson cluster process is utilized to generate clusters, the process of which has been described in Section II. The time length of a packet is assumed to be 10 ms. It should be noted that resource allocation is out of the research scope of this paper. We focus on enhancing the multicast performance by designing a proper multicast transmissions mechanism. Correspondingly, we employ the random network coding-based multicast as a compared scheme in this paper, which is a state-of-the-art approach and can significantly enhance the performance of multicast transmissions [11], [12]. The traditional multicast with ACK feedback and retransmissions is also adopted as a benchmark scheme for performance comparison, where a receiver needs to feed an ACK back for each received packet and the BS will retransmit the packet that have not been received by all receivers. The detailed simulation parameters are shown in Table I.

\subsection{Validation of results}

\vspace{-0em}\begin{figure}[t] 
  \begin{center}
    \scalebox{0.65}[0.65]{\includegraphics{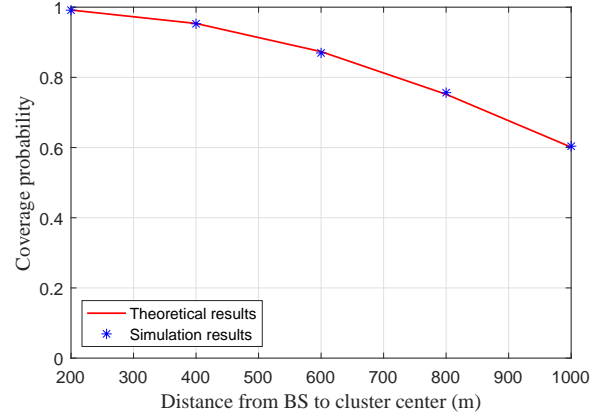}}
     \vspace{-0.0em}\caption{Coverage probability versus the distance from BS to cluster center $\left\| \mathbf{v} \right\|$.}
    \label{fig:1}
  \end{center}
\vspace{-0em}\end{figure}

\vspace{-0em}\begin{figure}[t] 
  \begin{center}
    \scalebox{0.6}[0.6]{\includegraphics{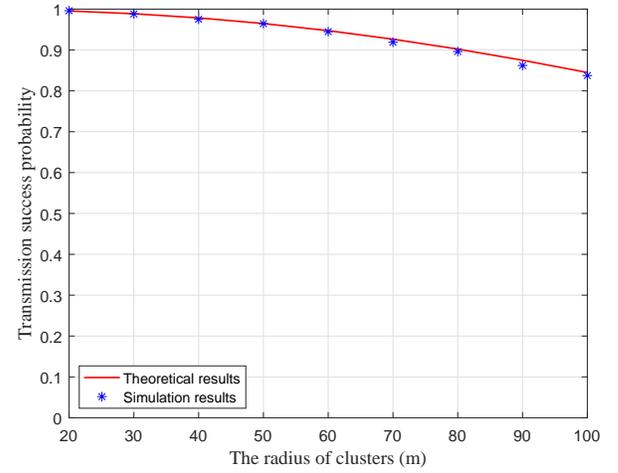}}
     \vspace{-0.6em}\caption{Transmission success probability versus the radius of clusters $r$.}
    \label{fig:1}
  \end{center}
\vspace{-0em}\end{figure}

The validation of the analytical performance analysis is investigated by comparing theoretical results and simulation results. Fig. 8 shows the results of coverage probabilities under the various distances from BS to cluster center $\left\| \mathbf{v} \right\|$, where the theoretical results are obtained based on the equations (6) and (8). In the simulation, 10000 UAVs are taken into account, which are uniformly distributed in a cluster with the radius of 50 $m$ in random. From Fig. 8, it can be observed that the coverage probability declines with the increase of $\left\| \mathbf{v} \right\|$. With a longer propagation, the SNR of received signals would be smaller, causing lower coverage probabilities. Fig. 9 presents transmission success probabilities under different cluster radiuses, in which the theoretical results are calculated according to the equations (7) and (9), while the simulation results are generated considering 10000 transmitter-receiver UAV pairs, each of which is randomly, independently, and uniformly distributed in a cluster. Apparently, the transmission success probability is shown to decrease with the growth of the radius, as with a larger radius the distance between UAVs may be relatively lengthened. Note that in both Fig. 8 and Fig. 9, the theoretical results perfectly match with the simulation results, which demonstrates the validation of the analytical performance evaluation in this paper.

\subsection{Design insight explorations}

\vspace{-0em}\begin{figure}[t] 
  \begin{center}
    \scalebox{0.6}[0.6]{\includegraphics{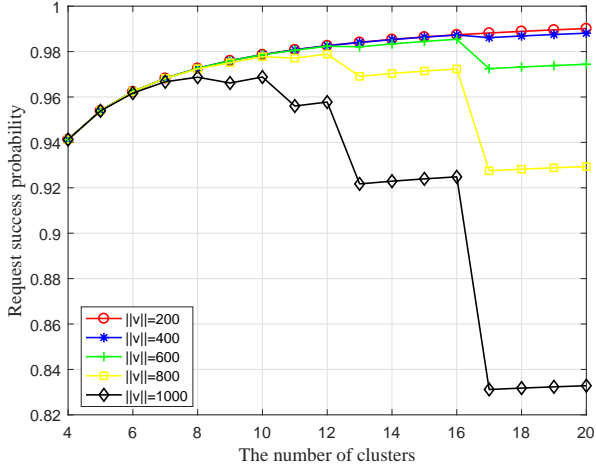}}     \vspace{-0em}\caption{Request success probability versus the number of clusters under various distances from BS to cluster center $\left\|  \mathbf{v} \right\|$.}
    \label{fig:1}
  \end{center}
\vspace{-0em}\end{figure}

\begin{figure*} [h] 
\centering
\subfigure[]{\label{fig:subfig:a}
\includegraphics[width=2.70in]{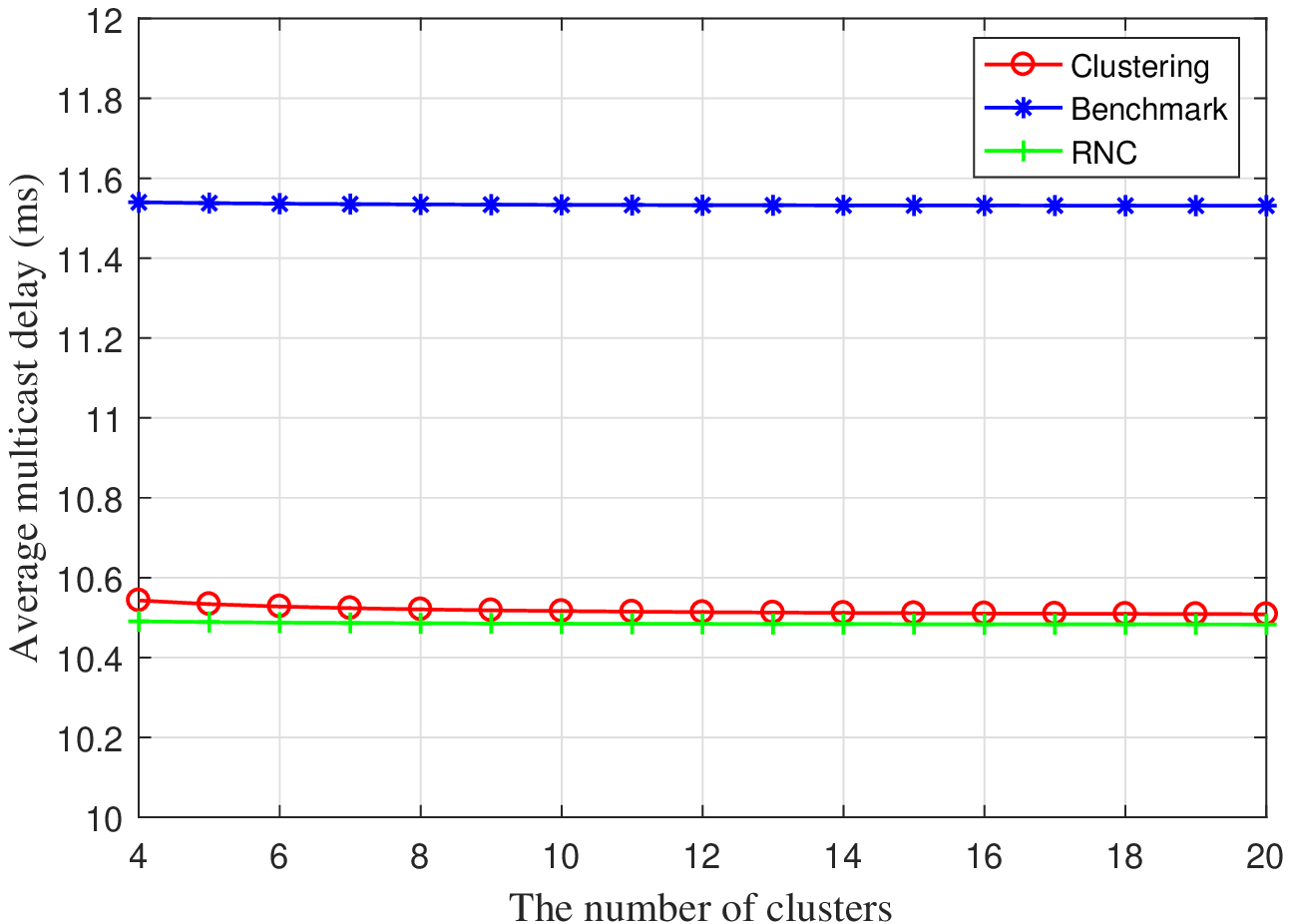}}
\hspace{-2.5em} \subfigure[]{ \label{fig:subfig:b}
\includegraphics[width=2.40in]{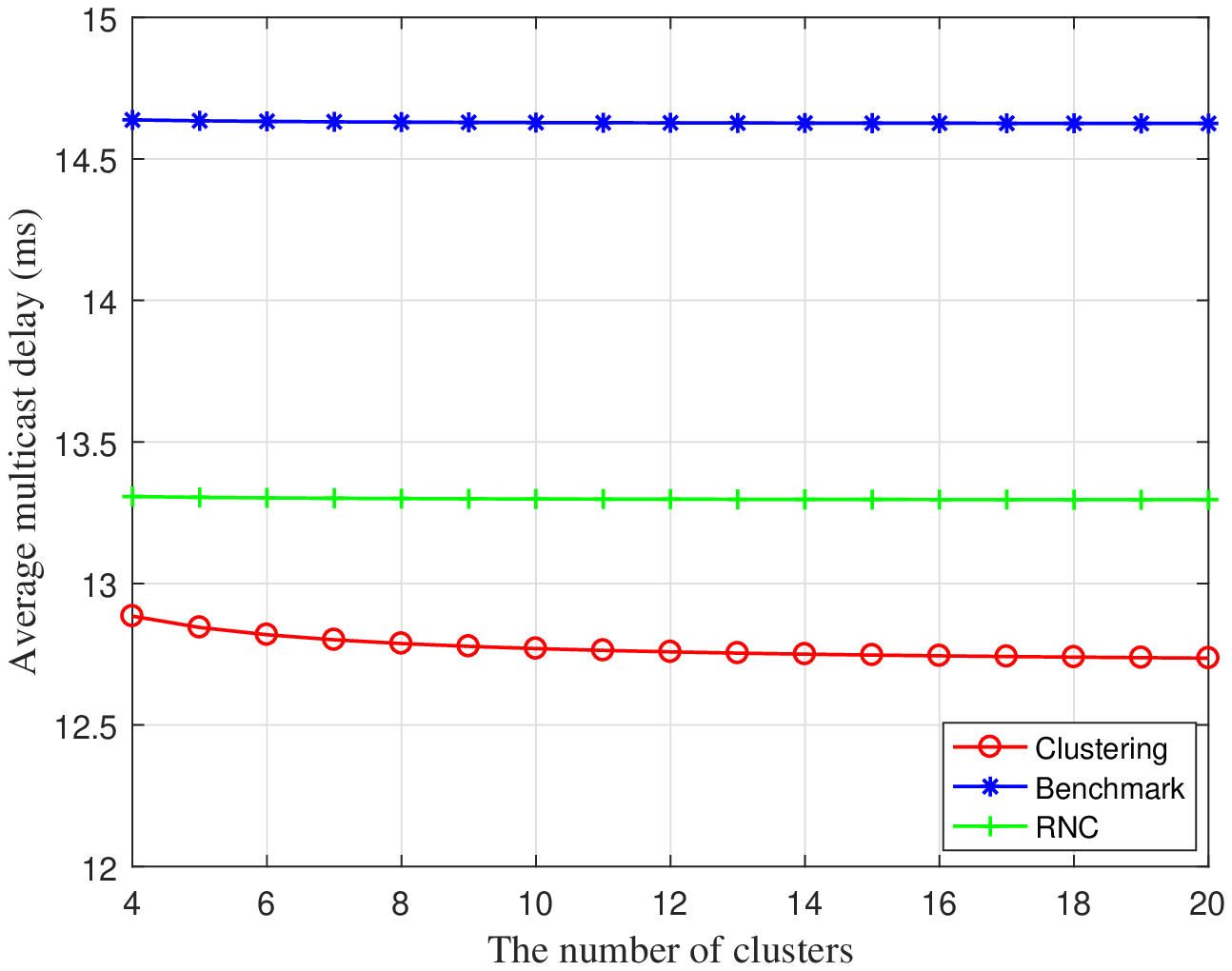}}
\hspace{-2.5em} \subfigure[]{ \label{fig:subfig:c}
\includegraphics[width=2.40in]{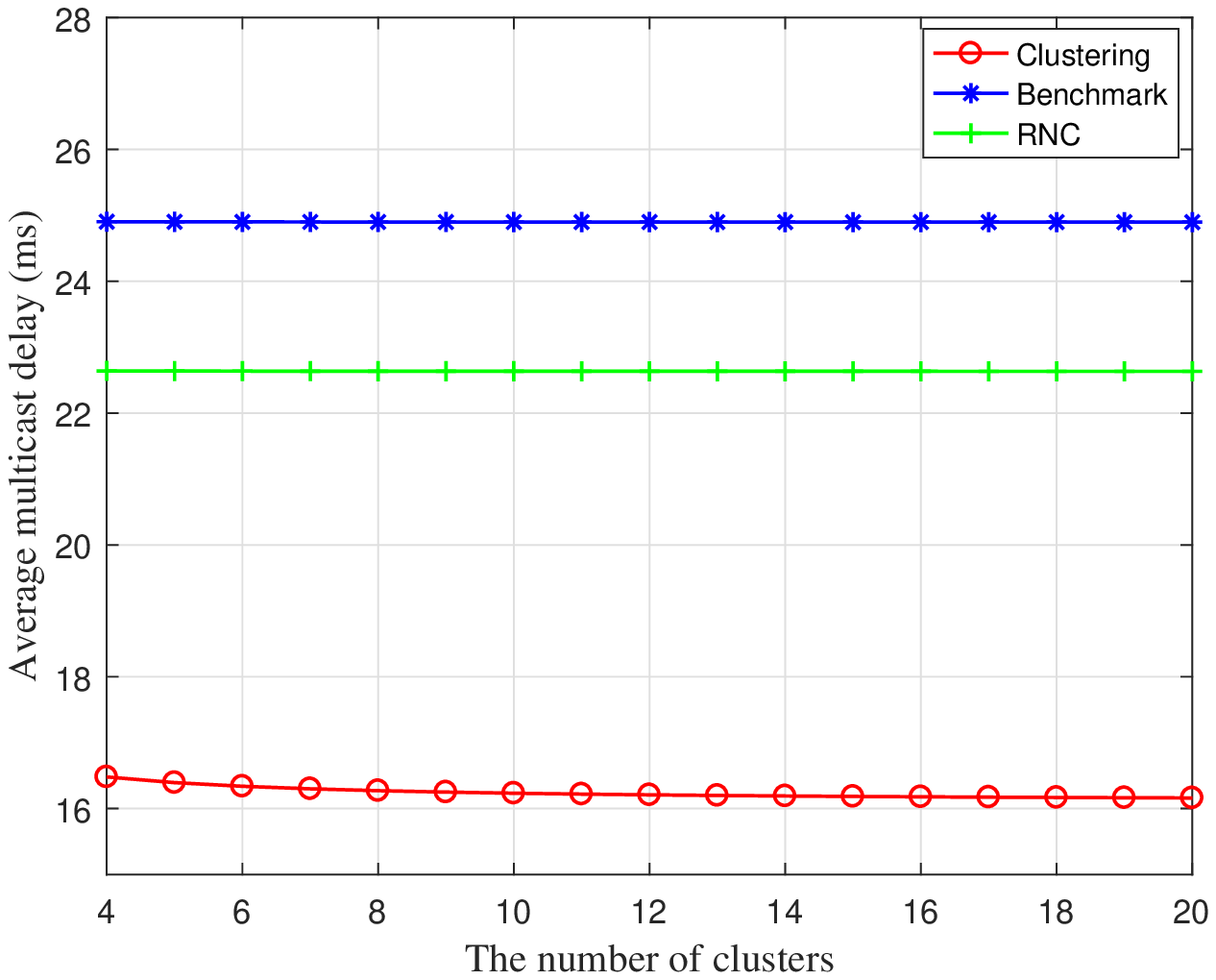}}
\hspace{-0em}\caption{Average multicast delay versus the number of clusters: a) ${D_0} = 400\;m$; b) ${D_0} = 800\;m$; c) ${D_0} = 1200\;m$.}
\label{fig:subfig}
\vspace{-0em}\end{figure*}

In Fig. 10, the request success probability in equation (10) is evaluated under different numbers of clusters $C$ and distances from BS to cluster center $\left\| \mathbf{v }\right\|$. In the simulation, a UAV network with 50 UAVs is considered with ${\lambda _{off}} = 1 \times {10^{ - 3}}/{m^2}$. Two main phenomena can easily be seen in Fig. 10. First, the request success probability significantly drops with the increase of $\left\| \mathbf{v } \right\|$. With a long propagation from a BS to a UAV network, the coverage probability would be low, meanwhile downgrading the request success probability. Second, with a small $\left\|  \mathbf{v} \right\|$, the request success probability will be improved with the growth of the cluster number $C$. This is because with a larger $C$, the UAV network will be partitioned into more clusters with a smaller cluster size, so that UAVs in a cluster are closer to each other, enhancing the transmission success probability. On the contrary, given a large $\left\|  \mathbf{v} \right\|$, the request success probability decreases with the increase of $C$. Clearly, if the cluster size is small with a large $C$, only a small number of UAVs exist in a cluster. Under the low coverage probability caused by a large $\left\| \mathbf{v} \right\|$, insufficient UAVs in a cluster will severely degrade the possibility that at least one UAV in the cluster can receive packets through BS's broadcasting.

According to these two phenomena, we could have a conclusion that when the distance from BS to cluster center $\left\| \mathbf{v} \right\|$ is short with an excellent coverage probability, the transmission success probability dominates the request success probability. Conversely, when $\left\| \mathbf{v} \right\|$ becomes large with a low coverage probability, the request success probability is mainly determined by the coverage probability. Accordingly, this conclusion could provide a system design insight that when a UAV network is deployed close to the BS, the UAV network should be partitioned into many clusters to restrict the cluster size and boost the transmission success probability. Contrarily, if a UAV network is far away from the BS, the number of clusters should be small to guarantee at least one UAV in a cluster able to receive broadcasted packets.

\subsection{Average broadcast delay}

Assume that the time for a UAV feeding an ACK back to BS and the time length of a packet request are both 1 $ms$. In the simulation, a UAV network is modeled with Poisson cluster process, where cluster centers are located using PPP in a circle area with the radius of 100 $m$ and 50 UAVs are uniformly distributed around them. Let ${D_0}$ be the distance between the BS and the center of the UAV network. Fig. 11 illustrates the average multicast delay of a packet under different ${D_0}$ and different amounts of clusters. It is obvious that owing to the property of RNC, the average delay of the RNC-based multicast is always lower than that of the traditional multicast method, the benchmark. Moreover, our designed multicast scheme using clustering significantly outperforms the RNC-based multicast and the benchmark on the average delay when the UAV network is far away from the BS, (${D_0} = 800\;m$ and ${D_0} = 1200\;m$). However, the average delay of our designed scheme is similar to that of the RNC-based multicast when the UAV network is close to the BS, ${D_0} = 400\;m$. The reason is that a small $D_0$ incurs high coverage probabilities and most of the UAVs can receive packets through BS's broadcasting with less packet loss. Whereas, when $D_0$ becomes large with a low coverage probability becomes low, the RNC-based multicast and the benchmark have to recover lost packets through BS's retransmissions, which is unreliable with a large $D_0$, resulting in severe multicast delay. With our designed scheme, lost packets are attained by short-range communications between UAVs in a cluster, which would be more reliable and efficient. It is noticeable that the cluster numbers have the trivial effect on the average multicast delay of our designed scheme. Even with the small number of clusters, making the cluster size relatively large, a UAV still communicates with another UAV in a short range with the high transmission success probability.

\subsection{Average area spectral efficiency}

\vspace{-0em}\begin{figure}[t] 
  \begin{center}
    \scalebox{0.6}[0.6]{\includegraphics{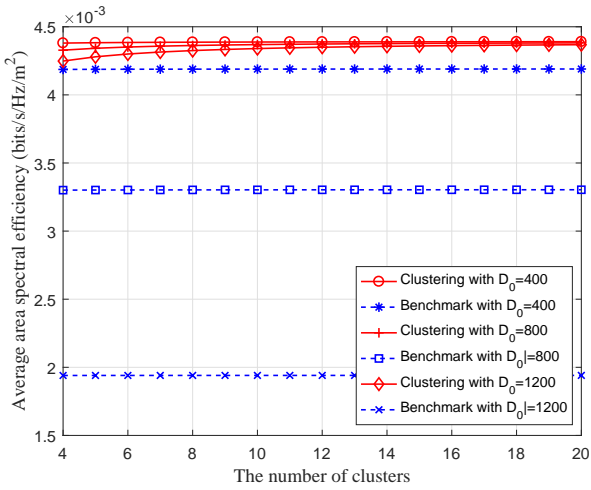}}
     \vspace{-0.0em}\caption{Average area spectrum efficiency versus the number of clusters under different $D_0$.}
    \label{fig:1}
  \end{center}
\vspace{-0.0em}\end{figure}

\vspace{-0em}\begin{figure}[t] 
  \begin{center}
    \scalebox{0.6}[0.6]{\includegraphics{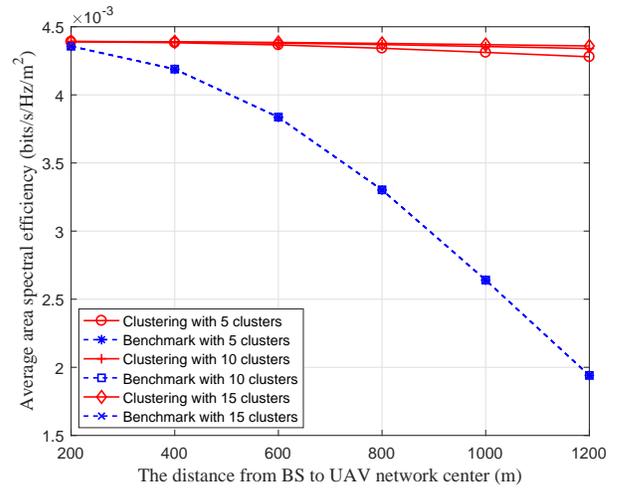}}
     \vspace{-0em}\caption{ Average area spectrum efficiency versus $D_0$ under different clusters numbers.}
    \label{fig:1}
  \end{center}
\vspace{-0em}\end{figure}

Fig. 12 and Fig. 13 plot the average area spectrum efficiency as a function of the number of clusters and $D_0$, respectively. As both the benchmark method and the RNC-based multicast method carry out multicast only depending on the broadcasting of BS, they share the same average area spectrum efficiency. In Fig. 12 and Fig. 13, an obvious trend is shown that our designed multicast scheme always surpasses the benchmark on the average area spectrum efficiency. With a larger $D_0$, more severe propagation loss will be encountered in the broadcasting from BS to UAVs, resulting in low coverage probabilities. For the benchmark method only relying on the broadcasting, its performance downgrades dramatically with the growth of $D_0$. In the multicast scheme with clustering, UAVs that experience packet loss regain the lost packets through transmissions between UAVs in the same cluster, which could always provide reliable and efficient packet delivery due to short-range transmissions. Thus, the average area spectrum efficiency can be preserved in a high level regardless of $D_0$. In addition, the number of clusters also has the slight influence on the proposed scheme, as no matter the changes of cluster amounts, transmissions between UAVs generally are performed in a relatively short range with high transmission success probabilities.

\section{Conclusions}

In this paper, a multicast scheme based on clustering is designed, where a UAV network will be partitioned into multiple clusters. If a UAV encounters a packet loss, it will request the lost packet from other UAVs in the same cluster rather than obtaining the lost packet by BS's retransmissions. Due to the fact that communications between UAVs are normally short-range wireless transmissions, the recovery of lost packets through transmissions between UAVs in a cluster would be more efficient compared to retransmissions from the BS. To guarantee practicability and without loss of generality, technical details of the introduced multicast scheme are sophisticatedly designed based on the CSMA/CA protocol. Through proper system design, some technical issues, such as signaling and packet storm, could be solved. Afterwards, the Poisson cluster process is adopted to model UAV networks to capture the dynamic network topology of UAV networks. Additionally, for analytical tractability, stochastic geometry is used to derive the distance distributions, with which the performance of the designed multicast scheme is studied analytically. Finally, by extensive simulation studies, we explore the system design insight of the optimal number of clusters. Moreover, the simulation results demonstrate the validation of the analytical analysis in this paper and show that our designed multicast scheme is able to support efficient multicast transmissions with a higher multicast delay and a higher area spectrum efficiency.

\section*{Appendix A}
\vspace{-0.0em}\section*{Derivations of equation (5)}

Recall that ${\hat d_1} = \sqrt {z_1^2 + z_2^2}$ and ${f_{{Z_1},{Z_2}}}\left( {{z_1},{z_2}\left| \mathbf{v} \right.} \right) = \frac{1}{{\pi {r^2}}}$, ${\left( {{z_1} - {v_1}} \right)^2} + {\left( {{z_2} - {v_2}} \right)^2} \leqslant {r^2}$. According to Fig. 14 and the law of Cosines, we have $\cos \,\alpha  = \frac{{\hat d_1^2 + {{\left\| \mathbf{v} \right\|}^2} - {{\left\| {{\mathbf{x}_0}} \right\|}^2}}}{{2{{\hat d}_1} \cdot \left\| \mathbf{v} \right\|}}$. If a UAV network is deployed far away from the BS working as an aerial small BS, $\hat d_1^2 >  > \left\| {{\mathbf{x}_0}} \right\|$ and $\left\| \mathbf{v} \right\| >  > \left\| {{\mathbf{x}_0}} \right\|$.
As a result, $\cos \,\alpha  = \frac{{ - {z_1}}}{{{{\hat d}_1}}} = \frac{{\hat d_1^2 + {{\left\| \mathbf{v} \right\|}^2} - {{\left\| {{\mathbf{x}_0}} \right\|}^2}}}{{2{{\hat d}_1} \cdot \left\| \mathbf{v }\right\|}}$ and $ {z_1} = - \frac{{\hat d_1^2 + {{\left\| \mathbf{v} \right\|}^2} - {{\left\| {{\mathbf{x}_0}} \right\|}^2}}}{{2\left\| \mathbf{v} \right\|}}$.
Given a $\mathbf{v}$ and a ${\hat d_1}$, the typical UAV, $\left( {{0},{0}} \right)$, can only lie on the dashed curve with blue color in Fig. 14, any point on which is ${\hat d_1}$ away from $\left( {{z_1},{z_2}} \right)$. Accordingly, $z_1$ would be maximum when $ \left\| {{\mathbf{x}_0}} \right\| = r$, while it is minimum when the typical UAV is located on the line between $\left( {{z_1},{z_2}} \right)$ and $\left( {{x_1},{x_2}} \right)$ with $\alpha = 0$. Thus, we have $ - {\hat d_1} \leqslant {z_1} \leqslant  - \frac{{\hat d_1^2 + {{\left\| \mathbf{v} \right\|}^2} - {r^2}}}{{2\left\| \mathbf{v} \right\|}}$.

\vspace{-0em}\begin{figure}[t] 
  \begin{center}
    \scalebox{1.0}[1.0]{\includegraphics{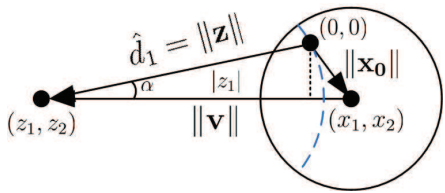}}
     \vspace{-0.0em}\caption{The distance from a BS to a typical UAV.}
    \label{fig:1}
  \end{center}
\vspace{-0.0em}\end{figure}

With ${\hat d_1} = \sqrt {z_1^2 + z_2^2}$, the conditional PDF of ${\hat d_1}$ conditioned on $\mathbf{v}$ could be derived as:

\vspace{-0em}\begin{align*}%
\footnotesize
\begin{array}{l}
\begin{gathered}
  {f_{{{\hat D}_1}}}\left( {{{\hat d}_1}\left| \mathbf{v} \right.} \right) = \int_{ - {{\hat d}_1}}^{ - \frac{{\hat d_1^2 + {{\left\| \mathbf{v} \right\|}^2} - {r^2}}}{{2\left\| \mathbf{v} \right\|}}} {{f_{{Z_1},{Z_2}}}\left( {{z_1},\sqrt {\hat d_1^2 - z_1^2} \left| \mathbf{v} \right.} \right)}
   \hfill \\
   \cdot\left| {\frac{{\partial \sqrt {\hat d_1^2 - z_1^2} }}{{\partial {{\hat d}_1}}}} \right|
   + {f_{{Z_1},{Z_2}}}\left( {{z_1}, - \sqrt {\hat d_1^2 - z_1^2} \left| \mathbf{v} \right.} \right)\left| { - \frac{{\partial \sqrt {\hat d_1^2 - z_1^2} }}{{\partial {{\hat d}_1}}}} \right|\;\mathrm{d}{z_1}
   \hfill \\
   = \frac{{2\hat d_1}}{{\pi {r^2}}}\int_{ - {{\hat d}_1}}^{ - \frac{{\hat d_1^2 + {{\left\| \mathbf{v} \right\|}^2} - {r^2}}}{{2\left\| \mathbf{v} \right\|}}} {\frac{1}{{\sqrt {\hat d_1^2 - z_1^2} }}} \;\mathrm{d}{z_1}
   \hfill \\
\end{gathered}
 \end{array}
\end{align*}

With the substitution of ${z_1} = {\hat d_1}\sin \,\theta $, we have:

\vspace{-0em}\begin{align*}%
\footnotesize
\begin{array}{l}
\begin{gathered}
  {f_{{{\hat D}_1}}}\left( {{{\hat d}_1}\left| \mathbf{v} \right.} \right) = \frac{{2\hat d_1}}{{\pi {r^2}}}\int_{ - \frac{\pi }{2}}^{ - \arcsin \,\frac{{\hat d_1^2 + {{\left\| \mathbf{v} \right\|}^2} - {r^2}}}{{2\left\| \mathbf{v} \right\|{{\hat d}_1}}}} {\frac{1}{{\sqrt {\hat d_1^2 - \hat d_1^2{{\sin }^2}\theta } }}} \;\mathrm{d}{{\hat d}_1}\sin \,\theta  \hfill \\
   = \frac{{2\hat d_1}}{{\pi {r^2}}}\left( {\frac{\pi }{2} - \arcsin \,\frac{{\hat d_1^2 + {{\left\| \mathbf{v} \right\|}^2} - {r^2}}}{{2\left\| \mathbf{v} \right\|{{\hat d}_1}}}} \right),\quad \left\| \mathbf{v} \right\| - r \leqslant {{\hat d}_1} \leqslant \left\| \mathbf{v} \right\| + r. \hfill \\
\end{gathered}
 \end{array}
\end{align*}

It is clear to see that the conditional PDF of ${\hat d_1}$ only needs to condition on $\left\| \mathbf{v} \right\|$ rather than $\mathbf{v}$.

\section*{Appendix B}
\vspace{-0em}\section*{Derivations of equation (7)}

From Fig. 15, $\cos \,\beta  = \frac{{{y_1}}}{{{d_2}}} = \frac{{d_2^2 + {{\left\| {{\mathbf{x}_0}} \right\|}^2} - {{\left\| \mathbf{w} \right\|}^2}}}{{2{d_2} \cdot \left\| {{\mathbf{x}_0}} \right\|}}$ and ${y_1} = \frac{{d_2^2 + {{\left\| {{\mathbf{x}_0}} \right\|}^2} - {{\left\| \mathbf{w} \right\|}^2}}}{{2\left\| {{\mathbf{x}_0}} \right\|}}$. It is important to note that when $d_2^2 + {\left\| {{\mathbf{x}_0}} \right\|^2} < {\left\| \mathbf{w} \right\|^2}$, $\beta $ is an obtuse angle and $y_1$ is a negative value. Conditioned on $\mathbf{x}_0$, $\left\| {{\mathbf{x}_0}} \right\|$ is a fixed value. With the span of $\left\| \mathbf{w} \right\|$ from 0 to $r$, $\frac{{d_2^2 + {{\left\| {{\mathbf{x}_0}} \right\|}^2} - {r^2}}}{{2\left\| {{\mathbf{x}_0}} \right\|}}$ is the minimum value of $y_1$, while $y_1$ is maximum when $\beta = 0$. Thus, we have $\frac{{d_2^2 + {{\left\| {{\mathbf{x}_0}} \right\|}^2} - {r^2}}}{{2\left\| {{\mathbf{x}_0}} \right\|}} \leqslant {y_1} \leqslant {d_2}$.


\vspace{-0em}\begin{figure}[t] 
  \begin{center}
    \scalebox{1.0}[1.0]{\includegraphics{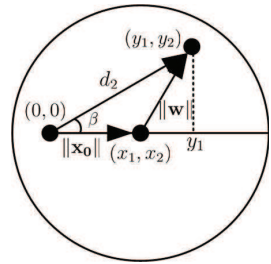}}
     \vspace{-0.0em}\caption{The distance between two UAVs.}
    \label{fig:1}
  \end{center}
\vspace{-0em}\end{figure}

The rest of the derivation of equation (7) is similar to that of equation (5). Please refer to the Appendix A.

\section*{Appendix C}
\vspace{-0em}\section*{Derivations of ${f_A}\left( a \right)$}

Recall that ${f_{{X_1},{X_2}}}\left( {{x_1},{x_2}} \right) = \frac{1}{{\pi {r^2}}}$, $x_1^2 + x_2^2 \leqslant {r^2}$. Let $A = \sqrt {x_1^2 + x_2^2} $ and $\Psi  = \arctan \frac{{{x_2}}}{{{x_1}}}$ be two random variables with respect to $x_1$ and $x_2$. The joint PDF of $A$ and $\Psi $ is:

\vspace{-0em}\begin{align*}%
\small
\begin{array}{l}
\begin{gathered}
  {f_{A,\Psi }}\left( {a,\phi } \right) = {f_{{X_1},{X_2}}}\left( {a\cos \phi ,a\sin \phi } \right)  \left| {\det \left[ {\begin{array}{*{20}{c}}
  {\frac{{\partial a\cos \phi }}{{\partial a}}}&{\frac{{\partial a\cos \phi }}{{\partial \phi }}} \\
  {\frac{{\partial a\sin \phi }}{{\partial a}}}&{\frac{{\partial a\sin \phi }}{{\partial \phi }}}
\end{array}} \right]} \right|
\hfill \\
   + {f_{{X_1},{X_2}}}\left( { - a\cos \phi , - a\sin \phi } \right)
    \left| {\det \left[ {\begin{array}{*{20}{c}}
  { - \frac{{\partial a\cos \phi }}{{\partial a}}}&{ - \frac{{\partial a\cos \phi }}{{\partial \phi }}} \\
  { - \frac{{\partial a\sin \phi }}{{\partial a}}}&{ - \frac{{\partial a\sin \phi }}{{\partial \phi }}}
\end{array}} \right]} \right| \hfill \\
   = \frac{{2a}}{{\pi {r^2}}}. \hfill \\
\end{gathered}
 \end{array}
\end{align*}

Obviously, $A$ and $\Psi$ are independent of each other according to ${f_{A,\Psi }}\left( {a,\phi } \right) = \frac{{2a}}{{\pi {r^2}}}$. The marginal PDF of $\Psi$ is ${f_\Psi }\left( \phi  \right) = \int_0^r {\frac{{2a}}{{\pi {r^2}}}} \;da = \frac{1}{\pi }$. Thus, the PDF of $A$ could be given by:

\vspace{-0.0em}\begin{align*}%
\small
\begin{array}{l}
{f_A}\left( a \right) = \frac{{{f_{A,\Psi }}\left( {a,\phi } \right)}}{{{f_\Psi }\left( \phi  \right)}} = \frac{{2a}}{{{r^2}}},\quad 0 \leqslant a \leqslant r.
 \end{array}
\end{align*}

\end{document}